\documentclass[twocolumn,prb,superscriptaddress]{revtex4-2}
\usepackage{amsmath,amsfonts,amssymb}
\usepackage{graphicx,color,xcolor}
\usepackage{hyperref}
\usepackage{epstopdf}
\usepackage{float}
\usepackage{multirow}
\usepackage{hhline}
\usepackage{ulem}
\usepackage{mathtools}
\usepackage{siunitx}
\usepackage[caption=false]{subfig}
\usepackage{soul}
\hypersetup{hidelinks,colorlinks=true,allcolors=BLUE}
\usepackage{blkarray}
\usepackage{amsmath}

\newcommand{\ket}[1]{\left|#1\right\rangle}

\makeatletter

\@ifundefined{textcolor}{}
{%
 \definecolor{BLACK}{gray}{0}
 \definecolor{WHITE}{gray}{1}
 \definecolor{RED}{rgb}{1,0,0}
 \definecolor{GREEN}{rgb}{0,1,0}
 \definecolor{BLUE}{rgb}{0,0,1}
 \definecolor{CYAN}{cmyk}{1,0,0,0}
 \definecolor{MAGENTA}{cmyk}{0,1,0,0}
 \definecolor{YELLOW}{cmyk}{0,0,1,0}
}

\makeatother

\begin{document}

\title{A Flux-Tunable Cavity for Dark Matter Detection}

\author{Fang Zhao}
\altaffiliation{These authors contributed equally to this work}
\affiliation{Fermi National Accelerator Laboratory, Batavia, IL 60510, USA}

\author{Ziqian Li}
\altaffiliation{These authors contributed equally to this work}
\affiliation{James Franck Institute, University of Chicago, Chicago, Illinois 60637, USA}
\affiliation{Department of Physics, University of Chicago, Chicago, Illinois 60637, USA}
\affiliation{Department of Applied Physics, Stanford University, Stanford, California 94305, USA}

\author{Akash V. Dixit}
\affiliation{James Franck Institute, University of Chicago, Chicago, Illinois 60637, USA}
\affiliation{Department of Physics, University of Chicago, Chicago, Illinois 60637, USA}
\affiliation{Kavli Institute for Cosmological Physics, University of Chicago, Chicago, Illinois 60637, USA}

\author{Tanay Roy}
\affiliation{James Franck Institute, University of Chicago, Chicago, Illinois 60637, USA}
\affiliation{Department of Physics, University of Chicago, Chicago, Illinois 60637, USA}

\author{Andrei Vrajitoarea}
\affiliation{Center for Quantum Information Physics, Department of Physics, New York University, New York 10003, USA}

\author{Riju Banerjee}
\affiliation{Department of Physics, University of Chicago, Chicago, Illinois 60637, USA}

\author{Alexander Anferov}
\affiliation{Department of Physics, University of Chicago, Chicago, Illinois 60637, USA}
\affiliation{Pritzker School of Molecular Engineering, University of Chicago, Chicago, Illinois 60637, USA}
\author{Kan-Heng Lee}
\affiliation{Department of Physics, University of Chicago, Chicago, Illinois 60637, USA}
\affiliation{Pritzker School of Molecular Engineering, University of Chicago, Chicago, Illinois 60637, USA}

\author{David I. Schuster}
\affiliation{James Franck Institute, University of Chicago, Chicago, Illinois 60637, USA}
\affiliation{Department of Physics, University of Chicago, Chicago, Illinois 60637, USA}
\affiliation{Department of Applied Physics, Stanford University, Stanford, California 94305, USA}
\affiliation{Pritzker School of Molecular Engineering, University of Chicago, Chicago, Illinois 60637, USA}

\author{Aaron Chou}
\affiliation{Fermi National Accelerator Laboratory, Batavia, IL 60510, USA}

\date{\today}

\begin{abstract}

Developing a dark matter detector with wide mass tunability is an immensely
desirable property, yet it is challenging due to maintaining strong sensitivity. Resonant cavities for dark matter detection have traditionally employed mechanical tuning, moving parts around to change electromagnetic boundary
conditions. However, these cavities have proven challenging to operate in sub-Kelvin cryogenic environments due to differential thermal contraction, low heat capacities, and low
thermal conductivities. Instead, we develop an electronically tunable cavity architecture by coupling a superconducting 3D microwave cavity with a DC flux tunable SQUID. With a flux delivery system engineered to maintain high coherence in the cavity, we perform a hidden-photon dark matter search below the quantum-limited threshold. A microwave photon counting technique is employed through repeated quantum non-demolition measurements using a transmon qubit. With this device, we perform a hidden-photon search and constrain the kinetic mixing angle to ${\varepsilon}< 8.2\times 10^{-15}$ in a tunable band from 5.672 GHz to 5.694 GHz. By coupling multimode tunable cavities to the transmon, wider hidden-photon searching ranges are possible.

\end{abstract}

\maketitle
\section{Introduction}
Dark matter (DM) is an integral part of the standard model of cosmology, though its specific composition is still a mystery.  If it comprises of particles with mass less than $\sim$100 eV, the inferred galactic DM mass density implies that the mode occupation number must be large~\cite{Randall_2017}.  In this mass range, DM may be modeled as classical, coherent waves that can exert weak, sinusoidal forces on laboratory oscillators with signal frequency $f$. Since dark matter is non-relativistic, the signal frequency is proportional to its rest mass $m$\cite{Sikivie1983}.  The strength of these forces is unknown;  while it must be weaker than electromagnetism because the DM is not luminous, since dark matter has only been observed via its gravitational imprint on cosmology and astrophysics, the couplings to mass-energy could be as weak as gravity~\cite{PhysRevD.102.072003} and still reproduce all cosmological observations.  A large range of allowed coupling strengths remains unexplored between these two limits.

Direct detection experiments perform radio searches for wavelike DM by tuning a laboratory oscillator over a range of resonant frequencies to search for excess noise power from DM disturbances above that estimated from data in neighboring frequency tunings~\cite{Graham:2015ouw}. The detection bandwidth $b=10^{-6} \ f$ is determined by the kinetic energy spread of the distribution of DM trapped in the galaxy's gravitational potential well with maximum velocity limited by the escape velocity~\cite{Baxter2021}.  In the small signal limit, individual signal photons in the oscillator can be detected using single-photon counters. Because these counters are insensitive to the phase of the signal oscillation, they can evade the standard quantum limit and reach noise levels limited only by Poisson fluctuations of signal and background counts.

With standard assumptions about the local DM density $\rho$~\cite{Baxter2021}, a signal excess or non-excess at a particular frequency tuning can be interpreted as a detection or constraint on the DM coupling to photons.
For DM in the form of dark or hidden-photons~\cite{Graham:2015rva,Chaudhuri:2014dla}, the driving force can be modeled as an effective current, which is given by
\begin{align}
j_{\text{HP}} &= \varepsilon m_{\gamma'} \sqrt{2\rho} e^{i m_{\gamma'} t} \hat{u}.
\end{align}
Here, $\varepsilon$ is the postulated kinetic mixing angle between standard electromagnetism and hidden sector electromagnetism, which determines the strength of the coupling and the hidden electric field amplitude of the DM wave is $E' = \sqrt{2\rho}$. Here, the polarization of the hidden-photon field is denoted by $\hat{u}$, and the mass of the hidden-photon is represented by $m_{\gamma'}$. A microwave cavity with a resonance frequency $f$ tuned to the hidden-photon mass $m_{\gamma'} = hf$ is used to coherently accumulate the electromagnetic response to the extremely undercoupled DM driving force. Here, $h$ is the Planck constant. Because the DM mass is unknown, it is critical to build detectors with wide frequency tunability.

Quantum techniques have brought unprecedented sensitivities in single microwave photon detection for DM searches\cite{PhysRevLett.126.141302, PhysRevLett.132.140801, PhysRevApplied.21.014043, braggio2024quantumenhanced, PhysRevLett.131.211001,chen2024searchqcdaxiondark}. 
Recently, the repeated quantum non-demolition measurements of fixed-frequency cavity photons through a superconducting qubit achieved a dark count rate of around 1/s
% with associated Poisson noise far below the standard quantum limit (SQL)
~\cite{PhysRevLett.126.141302}. The qubit-cavity interaction can also prepare non-Gaussian initial quantum states, which enhance the signal photon rate via stimulated emission~\cite{PhysRevLett.132.140801}.  Balembois et al. present an itinerant single microwave photon receiver based on an irreversible four-wave mixing process with a dark count rate 85 $s^{-1}$ and 50 MHz tunability\cite{PhysRevApplied.21.014043,braggio2024quantumenhanced}. DM searches are also possible by looking for the forces exerted directly on a qubit whose frequency is tunable using a SQUID~\cite{PhysRevLett.131.211001,chen2024searchqcdaxiondark}. However, the conductive RF-shielding enclosure may strongly suppress the electromagnetic response with a spectrum of modes at fixed frequencies.

Here, we use a DC flux-tunable SQUID to directly tune the frequency of a coupled cryogenic 3D microwave cavity. The flux delivery system is designed to maintain the cavity coherence required to accumulate and store severely undercoupled DM signals and to reject out-of-band noise. Unlike the tunable 2D resonators and tunable qubits concepts mentioned above, the 3D storage cavity offers a significantly larger volume, making it an ideal target for intercepting dark matter waves.  
Theoretically, the 3D cavity mode frequency can have a tuning range as large as the frequency separation in the avoided level crossing~\cite{Majer2007} between the storage and the coupler modes, approximately $2g \sim 200$ MHz, where $g$ is the coupling strength between storage and coupler modes. Operating multiple tunable cavities simultaneously expands the overall tuning range for DM experiments.

The structure of the paper is organized as follows: First, we present the theoretical framework and describe the experimental setup of the tunable photon counting detector. Second, we perform an experimental calibration of the detector's efficiency within the detection band. Finally, we analyze the results of the hidden-photon exclusion experiments and propose directions for future improvements.

\section{Tunable photon counting detector}

\begin{figure*}
    \centering
    \includegraphics[scale=0.9]{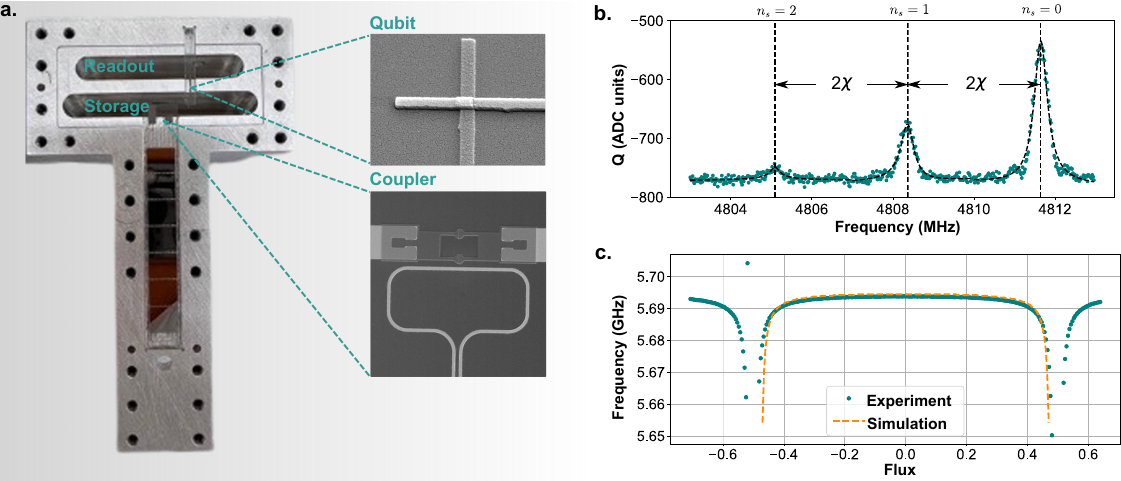}
    \caption{Tunable superconducting qubit hidden-photon detector. (a) The tunable hidden-photon detector comprises a ``Coffin'' cavity with two modes: readout (Top) and storage (bottom). The qubit chip (Top) is capacitively coupled to both modes, and the coupler chip (bottom) is coupled only to the storage mode for frequency tuning. The insets show the scanning electron micrograph of the qubit junction (Top) and the SQUID loop (Bottom). (b) Qubit frequency-resolved two-tone spectroscopy for a displaced state in the storage mode at $\Phi_{\text{ext}}=0$. In this measurement, the qubit is weakly driven while monitoring the readout; a response shift indicates the drive matches the qubit frequency modified by storage photons. Dash lines show the corresponding peak positions when the storage is at $\{\ket{0}, \ket{1}, \ket{2}\}$ respectively. (c) DC flux scan of storage mode single-tone spectroscopy. The storage mode peak positions are extracted and shown in the green dots. The Black-box simulation~\cite{PhysRevLett.108.240502} results are included in the yellow dash-line.}
    \centering
    \label{figure1}
\end{figure*}

Our tunable single-photon counter comprises four main components (see Fig.~\ref{figure1}(a)): a storage cavity; a coupler chip that includes both an on-chip flux filter and a DC SQUID; a qubit chip; and a readout cavity, which shares the same “Coffin” cavity as the storage. Each component hosts a single mode utilized in the experiment: A high-Q (Q$>10^6$) tunable storage mode $\omega_{s}$ that stores single-photon signals in the cavity~\cite{BARBIERI2017135}, a symmetric DC SQUID mode $\omega_{c}$ that tunes the storage mode frequency while maintaining cavity coherence, a dispersively coupled transmon $\omega_{q}$~\cite{PhysRevA.76.042319} and a readout mode $\omega_{r}$ which implements the quantum non-demolition (QND) measurement of the storage  photons~\cite{PhysRevLett.65.976, Gleyzes2007}. The system's Hamiltonian is

\begin{align}
H(\Phi_{\text{ext}}) &= \omega_s(\Phi_{\text{ext}}) a_s^{\dagger}a_s + \omega_q a_q^{\dagger}a_q + \omega_c(\Phi_{\text{ext}}) a_c^{\dagger}a_c + \omega_r a_r^{\dagger}a_r \nonumber \\
&+ \frac{\alpha_q}{2} a_q^{\dagger}a_q^{\dagger}a_q a_q + 2\chi(\Phi_{\text{ext}}) n_q n_s \nonumber \\
&+ 2\chi_{qr} n_q n_r + 2\chi_{sc}(\Phi_{\text{ext}}) n_s n_c,
\end{align}
Here, $a$ and $n$ are the annihilation and photon number operators with $s,q,c$, and $r$ denoting the storage, transmon, coupler, and readout modes, respectively. The transmon–storage, transmon–readout, and storage–coupler coupling strengths are denoted by $\{\chi, \chi_{qr}, \chi_{sc}\}$, respectively, and $\Phi_{\text{ext}}$ is the DC flux through the coupler SQUID loop. For simplicity, $\Phi_{\text{ext}}$ is expressed in units of the magnetic flux quantum $\Phi_0$. Fig.~\ref{figure1}(b) shows the storage mode photon-number resolved peaks with $\chi=-2\pi\times \SI{1.655}{\mega\hertz}$ when $\Phi_{\text{ext}}=0$. The photon-resolved $\chi$ is necessary to map storage mode signals to the transmon state efficiently, then dispersively read out through the transmon-readout coupling $\chi_{qr}=-2\pi\times \SI{0.14}{\mega\hertz}$. Here the linewidths $\{\kappa_q, \kappa_s\}$ for the qubit and storage are separately $\{4.7, 2.4\}$ kHz. The DC flux tunable coupler mode $\omega_c$ is capacitively coupled to the storage mode, which allows for tuning $\omega_{s}$ by biasing $\omega_{c}$. The theoretical tuning range for our detector can reach up to twice the strength of the capacitive coupling, which is on the order of $\SI{100}{\mega\hertz}$ (See Appendix II for detail~\cite{SuppMat}). The other stray couplings (coupler-transmon, coupler-readout) are negligible; Therefore, the transmon frequency, readout frequency, and transmon anharmonicity $\{\omega_{q}, \omega_{r}, \alpha_q\}=2\pi\times\{4811,6773,-176\}$ MHz are approximated as constants in theory. Fig.~\ref{figure1} (c) displays a comparison between black-box quantization~\cite{PhysRevLett.108.240502} and experimental spectroscopy data, showing a good match across the DC flux region ($\Phi_{\text{ext}}\in\left[-0.4793\pi, 0\right]$) where our scheme is implemented. This corresponds to a detector tuning range of $\SI{22.8}{\mega\hertz}$ in our experiment. The SQUID frequency limits the tuning range: in our case, the SQUID frequency is too high to see the full storage-coupler avoided crossing within a DC flux range that maintains a reasonable storage dephasing time.

\begin{figure}[h]
    \centering
  \includegraphics[scale=0.83]
  {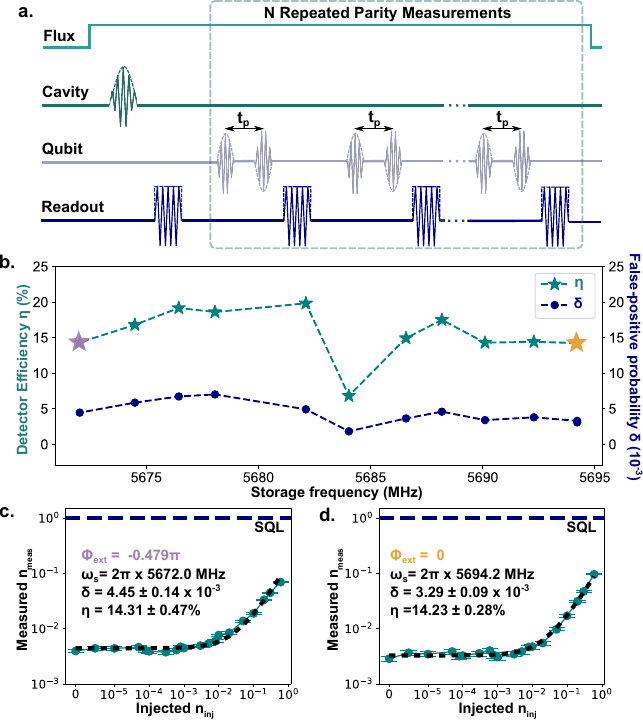}
    \caption{(a) Photon counting scheme. The pulse sequence at each DC flux bias point includes an initial weak storage displacement for detector efficiency characterization or a hidden-photon signal integration period. N repeated parity measurements are performed afterward, containing a readout pulse, readout photon releasing period, and two qubit $\pi/2$ rotations gapped by idling period $t_p=\pi/|2\chi|$. N bits of single shot readout after each detection cycle are collected. The probability of seeing false-positive events from the detector is exponentially suppressed in our detection scheme. (b) Detector efficiency characterization. At each flux biasing point, a list of pre-calibrated weak displacement drives is separately applied to the storage, followed by $N=25$ parity measurements and Markovian analysis with threshold $\lambda_{\text{thresh}}=125$. $\eta$ and $\delta$ are determined from the linear fitting. At $\Phi_{\text{ext}}=0$, the thermal population of the storage is quantified using the Fast Fourier Transformation (FFT) of the transmon Ramsey signal. Two detector characterizations ($\Phi_{\text{ext}}=-0.479\pi$ and $0$) are shown in (c). More characterization data are shown in Appendix IV~\cite{SuppMat}.}
    \centering
    % something else to show in the main text?
    \label{figure2}
\end{figure}

Fig.~\ref{figure2} (a) shows our photon counting scheme using Ramsey interferometry readout~\cite{PhysRevLett.126.141302}, which detects the parity of the storage mode photon population. For each new detection frequency, $\omega_{s}$ is DC flux biased according to the flux-frequency mapping shown in Fig.~\ref{figure1} (c). For flux point requiring detector efficiency calibration, we initiate the experiment by adding photons to the storage cavity. For actual hidden-photon detection, the experiments instead begin with a waiting period corresponding to the signal integration time. Following this procedure, N repeated parity measurements are performed: In each parity measurement, a readout tone ($\SI{2.3}{\micro\second}$ long) is first sent to check the current transmon state in $\ket{g}$ or $\ket{e}$. This is followed by a $\SI{5}{\micro\second}$ pause to allow for the readout to reset. Subsequently, a fast $\pi/2$ transmon rotation is applied using a $2\sigma$ tailed Gaussian pulse (where $\sigma=\SI{10}{\nano\second}\ll 1/\chi(\Phi_{\text{ext}})$), positioning the transmon into a superposition state $(\ket{g}\pm\ket{e})/\sqrt{2}$. The transmon then waits for $t_p=\pi/|2\chi|$ to acquire a phase that depends on the storage state. After another $\pi/2$ rotation, the transmon state is flipped if the storage mode is in an odd-number Fock state. A positive detection of a photon is indicated by observing a repeated sequence of transmon state transitions during the N parity measurements. In our detection protocol, we assume that the signals received by the storage mode are sufficiently weak, focusing only on the population within the first two levels of the storage mode. Therefore, the transmon state is flipped when the storage mode is excited.  

To resolve all possible error mechanisms during the measurement protocol, we model the evolution of the storage cavity, transmon, and readout as a hidden Markov process where the cavity and transmon states are hidden variables that emit as a readout signal~\cite{PhysRevLett.126.141302}, and the storage initial state probability $\{P_0, P_1\}$ is reconstructed using the backward algorithm. Here, $P_0$ and $P_1=1-P_0$ represent the calculated probabilities of the storage mode being in the states $\ket{0}$ and $\ket{1}$. In our experiment, we define a state discrimination threshold $\lambda_{\text{thresh}} = P_1/P_0$, and the false positive probability $\delta$ is exponentially suppressed as $\lambda_{\text{thresh}}$ increases. The storage mode is considered in $\ket{1}$ when $\lambda_{\text{thresh}} \geq 125$. Details about the hidden Markov process analysis and choosing $\lambda_{\text{thresh}}$ are discussed in Appendix IV~\cite{SuppMat}.

The storage and readout modes are machined on the same high-purity (6N) Aluminum cavity. Details about the cavity etching are discussed in Appendix I~\cite{SuppMat}. An edge-coupled microstrip on-chip filter is used on the flux-biasing line to protect the storage mode coherence. Details about the on-chip flux filter design are shown in Appendix II~\cite{SuppMat}.

\section{Detector efficiency characterization}

The tunable photon counting detector efficiency $\eta$ is characterized by initially injecting different predetermined populations of photons $\Bar{n}_{\text{inj}}$ into the storage mode through weak displacement drives. The weak storage displacement $\beta$ ($\beta\ll 1$) ensures the population remains mostly in the Fock states $\{\ket{0}, \ket{1}\}$. The calibration of the storage mode displacement is discussed in Appendix III~\cite{SuppMat}. The detector efficiency can then be determined through the proportion of positive events $\Bar{n}_{\text{meas}}$ after the Ramsey interferometry readout: We choose $N=25$ to discriminate between one and zero photon events in the storage mode. By varying $\Bar{n}_{\text{inj}}$, we fit this relationship with the function $\Bar{n}_{\text{meas}}=\eta\Bar{n}_{\text{inj}}+\delta$ to extract the detector efficiency $\eta$ and false positive probability $\delta$, defined as the background excitations in the storage detected by our photon counting scheme.

Fig.~\ref{figure2} (b) displays the detector characterization results across a range of flux tuning points. Two detailed detector efficiency plots at two flux tuning points are shown in Fig.~\ref{figure2}c. The other flux points' characterizations are in Appendix IV~\cite{SuppMat}. The detector efficiency remains reliably above $14\%$,  even when $\Phi_{\text{ext}}$ is close to $\pi/2$. This consistent performance is attributed to three key factors: First, the transmon and storage $T_1$ are insensitive to $\Phi_{\text{ext}}$, which are protected by the on-chip flux line filter as detailed in Appendix II~\cite{SuppMat}. Second, the tunable hidden-photon detector is resilient against transmon bit-flip error through continuous parity measurements. The hidden-photon Markov process analysis (see below) also compensates for variations in the storage decay time $T_{1}^s$ between tuning points.

The primary limitation in detector performance arises from storage cavity coherence and the background of real photons. We perform the master equation simulation to capture the detector performance (See Appendix V~\cite{SuppMat}) and show that the detector efficiency is limited to $25\%$ with our current setup's coherence. The storage coherence time limits $N$ before the stored photons decay. Spurious excitations of photons in the storage mode alter the parity readout pattern, leading to false positive events. $\delta$ increases as $\Phi_{\text{ext}}$ is tuned closer to $\pi/2$. We suspect that the enhanced hybridization between the coupler and storage modes near $\pi/2$ accelerates the transfer of the coupler's quasiparticle tunneling events into storage excitations. Additionally, the thermal population of the storage at $\Phi_{\text{ext}}=0$ is quantified using the Fast Fourier Transformation (FFT) of the transmon Ramsey signal. The measured intensity ratio at $\Phi_{\text{ext}}=0$ for transmon frequency at $\omega_{q}$ and $\omega_{q}-\chi$ is $8.7 \pm 0.5\times 10^{-3}$~\cite{PhysRevB.86.180504}. This measurement suggests that the storage thermal population $\Bar{n}_{c}=8.6 \pm 0.5 \times 10^{-3}$, corresponding to a cavity temperature of $57 \pm 1.4$ mK. We choose the detection threshold $\lambda_{\text{thresh}}=125$ for varying flux tuning points~\cite{PhysRevLett.126.141302} such that the background photon probability ($1/(\lambda_{\text{thresh}}+1) \textless \Bar{n}_c$) is comparable to the storage thermal population.

The measured storage temperature exceeds the mixing chamber temperature (15 mK), possibly because of quasiparticle tunneling events in the SQUID and the transmon being projected into stored photons via the dressing of the cavity\cite{PhysRevLett.126.141302}. Gap engineering~\cite{PhysRevLett.108.230509}, quasiparticle trapping~\cite{Pan2022}, and comprehensive infrared radiation shielding~\cite{Kreikebaum_2016} can effectively mitigate these effects.

\section{hidden-photon dark matter exclusion}

\begin{figure*}[t]
    \centering
    \includegraphics[width=2\columnwidth]{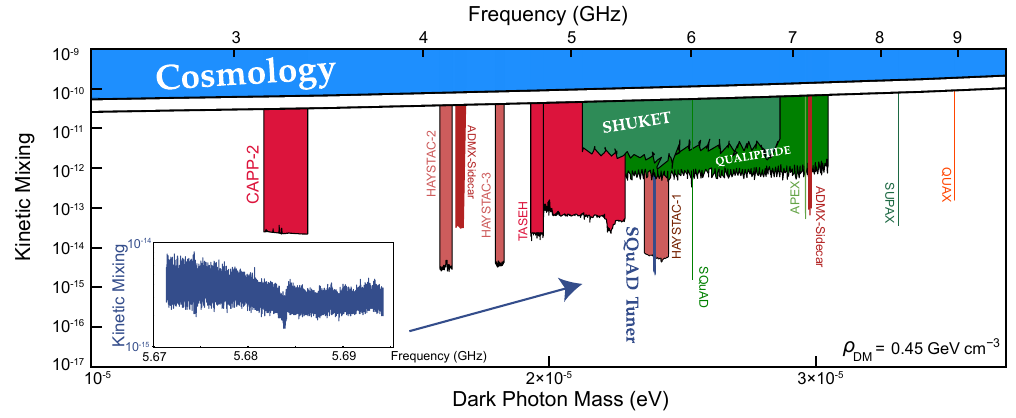}
    \caption{Comparison with previous dark photon exclusion results~\cite{Witte2020, PhysRevLett.125.221303, Paola_Arias_2012, PhysRevLett.122.201801, PhysRevLett.126.141302, PhysRevLett.130.231001, PhysRevLett.121.261302, PhysRevD.97.092001, Backes2021, haystaccollaboration2023new, PhysRevLett.125.221302, Adair2022, PhysRevLett.129.111802, PhysRevD.99.101101, schneemann2023results, PhysRevLett.107.191804}. Our exclusion data is labeled as ``SQuAD\_Tuner''. Inset: Hidden-photon exclusion results measured in this experiment. The storage cavity frequency is swept by $22.8$ MHz, and at each frequency point, 20,000 independent parity measurements are collected. The exclusion results at $95\%$ confidence level are analyzed based on Reference~\cite{PhysRevLett.126.141302}. All flux-dependent parameters are linearly interpolated and used in the analysis. The plot is generated using the code from Reference~\cite{PhysRevD.104.095029}. The geometric factor is not included in the plot for comparison.} 
    \centering
    \label{figure4}
\end{figure*}

By measuring photon counts through repeated parity assessments and utilizing a Markov model for analysis, we conduct a hidden-photon dark matter search over a 22.8 MHz range. We sample the tuning range with 12695 DC flux tuning points, achieving a frequency resolution of $2.8$ kHz for $83.7\%$ of the tuning steps, commensurate with half the dark matter linewidth ($5.7$ kHz) ~\cite{Turner1990}. Smaller step sizes could be achieved in future experiments by increasing the attenuation on the DC line inside the fridge. We collect 20,000 independent parity measurements at each tuning point. For each measurement, we perform $N=25$ Ramsey interferometry sequences with a readout length $t_r=2.3\,\mu$s, readout photon releasing time $t_l=5\,\mu$s, system reset time and signal integration time $t_c=500\,\mu$s. For detector characterization, the values of $t_c$ are chosen to be greater than five times the storage mode’s $T_1$ in our setup.  Within the same dataset, for the data with the calibration signal source turned off, the dark matter storage signal integration time $t_i$ is assumed to be only the storage lifetime $T_1^s\in\left[64.5, 69.2\right]\,\mu$s. Each measurement cycle time is thus $t_c+(t_r+t_l)\times25=682.5\,\mu$s with $182.5 \mu$s used for the readout sequence.   %with a duty cycle of $26.7\%$.} 
The system reset time is required during the calibration process; however, in the actual detection experiment, no reset is necessary, and the entire duration $t_c$ is dedicated to signal integration. The experiment has a duty cycle of $27\%$, with a measurement time of $10.65\,\mathrm{s}$ per flux tuning point. The $22.8\,\mathrm{MHz}$ scan is completed in $48.1\,\mathrm{hours}$, enabling future GHz-range tuning to finish within a reasonable amount of time.

We choose a detector threshold of $\lambda_{\rm thresh} = 125$ and count a maximum of 91 photons at 5680.2859 MHz as seen in Fig.~\ref{figure4}. To search for hidden-photon candidates in the tuning range [$5671.395$, $5694.212$] MHz, we first obtain a photon count background by smoothing the counts as a function of frequency using a Savitzky–Golay filter of order four and window length of 112 \cite{Yi2023}. We expect a dark matter signal to be distributed over two frequency bins and ensure the filter window size is much larger than the potential signal. We compute the cumulative probability of a dark matter candidate producing the observed signal above the measured background. This cumulative probability calculation takes the form $P(\leq N_{obs}) = 1-\sum_{N_{DM}=0}^{N_{obs}} \frac{(N_{back} + N_{DM})^{N_{obs}} e^{-(N_{back} + N_{DM})}}{N_{obs}!}$. Additionally, we account for the systematic uncertainties by marginalizing over the relevant measured parameters (See Appendix VI~\cite{SuppMat} for detail). Dark matter candidates that could produce a signal less than or equal to the observed counts with less that 5\% probability are excluded at the 95\% confidence level. See supplemental for details on determining and accounting for the background level and incorporating systematics into the 95\% confidence level exclusion of dark matter hypothesis. Figure~\ref{figure4} shows hidden-photon candidates that we exclude at the $95\%$ confidence level in the context of other experiments sensitive to wavelike dark matter.

\section{Conclusion}

In this paper, we demonstrate a state-of-the-art electronically tunable photon counter for dark matter detection, achieving an approximately 20-fold speedup compared to a linear quantum-limited amplifier~\cite{PhysRevD.88.035020, PhysRevLett.126.141302}. By coupling to a DC-tunable SQUID, we realized a tunability of 22.8 MHz in our storage cavity. Future devices should be able to use the level repulsion~\cite{Majer2007} at the avoided crossing to achieve the entire tuning range of $\sim 200$ MHz; the demonstration of SQUID-based tuning of a 3D cavity provides a promising avenue for future dark matter searches with GHz tuning range, employing many such tunable cavity modes, each with $3\%$ tuning range. The instantaneous tuning range can also be expanded by coupling a single SQUID to a multimode cavity system. This enables concurrent tuning of each cavity mode to search for the dark matter at each mode frequency simultaneously.

\section{ACKNOWLEDGEMENTS}
We would like to acknowledge and thank A. Agrawal, M. Lynn, and A. Oriani for discussions, and S. Uemura for RFsoc measurement supporting. This document was prepared using the resources of the Fermi National Accelerator Laboratory (Fermilab), a U.S. Department of Energy, Office of Science, Office of High Energy Physics (OHEP) User Facility, and supported by the OHEP's Quantum Information Science Enabled Discovery (QuantISED) program. Fermilab is managed by Fermi Forward Discovery Group (FFDG), acting under Contract No. DE-AC02-07CH11359.

\section{Data availability}
The data that support the findings of this article are openly available~\cite{li_2025_squad_tuner}.

\clearpage

\bibliography{2Q}
\end{document}

\title{Appendix for ``A Flux-Tunable Cavity for Dark Matter Detection"}

\maketitle

\section{Cavity and quantum devices fabrication}
\label{app:fab}
The cavities used in this work are fabricated from high-purity (99.9999\%) Aluminium. The full device consists of two microwave cavities (readout and storage), each coupled to the transmon. The storage cavity is coupled with one flux-tunable SQUID device.

The quantum devices were fabricated on 430 $\mu$m thick C-plane edge-fed growth Sapphire wafers with a diameter of 50.8 mm. Wafers were annealed in an oxygen-rich environment at 1200 for 3 hours. Afterwards, wafers were cleaned with three steps: 1. solvents clean(Toluene, Acetone, Methanol, Isopropanol, and DI water) in an ultrasonic bath; 2. Nanostrip clean at 45 C for 15 min and DI water rinse for 10min; 3, sulfuric acid clean at 70 C for 10min followed by a thorough rinse in DI water for 20min. Tantalum was deposited after the annealing and cleaning by DC magnet sputtering while maintaining a substrate temperature of 800 C. The features were defined via optical lithography using AZ MiR 1518 photoresist and exposure with a Heidelberg MLA150 Direct Writer. The resist was developed for 1 min in AZ MIF 300 1:1. The features were etched in a Plasma-Thermal inductive coupled plasma etcher using chlorine-based etch chemistry with a plasma consisting of 30 sccm Cl2, 30 sccm $BCl_3$ and 10 sccm Ar. After etching, the photoresist was removed in N-Methylpyrrolidone(NMP) overnight, followed by acetone, isopropanol, and DI water with sonicating for 5 min. 

Josephson junctions were patterned using electron-beam lithography. The wafer was first dehydrated by baking at 180 C for 8 mins under low vacuum. Then, a bilayer of MMA EL13 and 950K PMMA A7 was spun with a  5min bake at 180 C following the spinning of each layer. To eliminate charging effects during electron-beam writing, a 10 nm gold anti-charging layer was deposited by electron-beam evaporation. Electron-beam lithography was then performed using a Raith EBPG 5000+ to define the Manhattan pattern for transmon and the Dolan-bride for SQUID. The resist stack was developed for 90 seconds at 6 C in a solution of 3 parts IPA and 1 part DI water. The wafer was then loaded into the load lock of a Plassys UMS300 electron-beam evaporator. Before deposition, the overlap regions on the pre-deposited capacitors were milled in situ with an Argon ion mill to remove the native oxide. The junctions were then deposited using a three-step electron-beam evaporation and oxidation process. First, an initial 45 nm layer of aluminum was deposited at 1 nm/s. Next, the junctions were exposed to a high-purity mixture of Ar and O2 (ratio of 80:20) for the first layer to grow a native oxide. Finally, a second 115 nm layer of aluminum was deposited at 1 nm/s. After the second aluminum deposition, a second static oxidation step was performed using the same Ar:O2 mixture at 3 mbar for 5 mins in order to cap the surface of the bare aluminum with pure aluminum oxide. The wafer was then coated with a protective resist before dicing into individual chips. Liftoff was then performed by immersing the chips in NMP at 80 C for 3 hours followed by sonication for 2 mins each in NMP, acetone, and isopropanol. After the liftoff, the chips were exposed to an ion-producing fan for 15min to avoid electrostatic discharge of the junctions. The room temperature DC resistance of the Josephson junction on each transmon was measured to select the transmon that corresponds to the target Josephson energy~\cite{PhysRevA.76.042319}. 

The cavities underwent a chemical etching treatment using a mixture of phosphoric and nitric acid (Transene Aluminum Enchant Type A) heated to 50 C for etching aluminum 50 $\mu$m film. The transmon and coupler are inserted into the slots and clamped by the clamps on the other part of the cavities.

\section{Coupler and on-chip filter design}
\label{app:filter}
The Hamiltonian for the circuit containing the coupler and storage modes in the bare basis is described by

\begin{align}
H_{cs} = \tilde{\omega}_s\tilde{a}^{\dagger}_s\tilde{a_s}+\tilde{\omega}_c(\Phi_{\text{ext}})\tilde{a}^{\dagger}_c\tilde{a_c}+g\left(\tilde{a}^{\dagger}_s+\tilde{a_s}\right)\left(\tilde{a}^{\dagger}_c+\tilde{a_c}\right).
\end{align}

The set $\{\tilde{\omega}_s, \tilde{\omega}_c(\Phi_{\text{ext}})\}$ represents the bare basis frequencies of the storage mode and the coupler, respectively. The corresponding annihilation operators are denoted as $\{\tilde{a}_s, \tilde{a}_c\}$. The capacitive coupling strength is approximated by the constant $g$ within our detector's operating regime. By applying the rotating wave approximation (RWA), we derive the Jaynes-Cummings Hamiltonian

\begin{align}
H_{cs} = \tilde{\omega}_s\tilde{a}^{\dagger}_s\tilde{a_s}+\tilde{\omega}_c(\Phi_{\text{ext}})\tilde{a}^{\dagger}_c\tilde{a_c}+g\left(\tilde{a}^{\dagger}_s\tilde{a_c}+\tilde{a}^{\dagger}_c\tilde{a_s}\right).
\end{align}

Treating the coupler mode as a two-level system, storage mode's first excited state is

\begin{align}
\omega_s(\Phi_{\text{ext}})=\frac{\tilde{\omega}_s+\tilde{\omega}_c}{2}+\frac{1}{2}\sqrt{4g^2+\left(\tilde{\omega}_s-\tilde{\omega}_c(\Phi_{\text{ext}})\right)^2}.
\end{align}

The storage mode frequency is tunable via DC flux, allowing the full avoided-crossing frequency response to exhibit a maximum tuning range of \(2g\). In experimental setups, achieving a coupling strength greater than \(100\) MHz is straightforward, thereby theoretically enabling a maximum detector tuning range beyond \(200\) MHz.

\begin{figure}[t]
    \centering
    \includegraphics[width=\columnwidth]{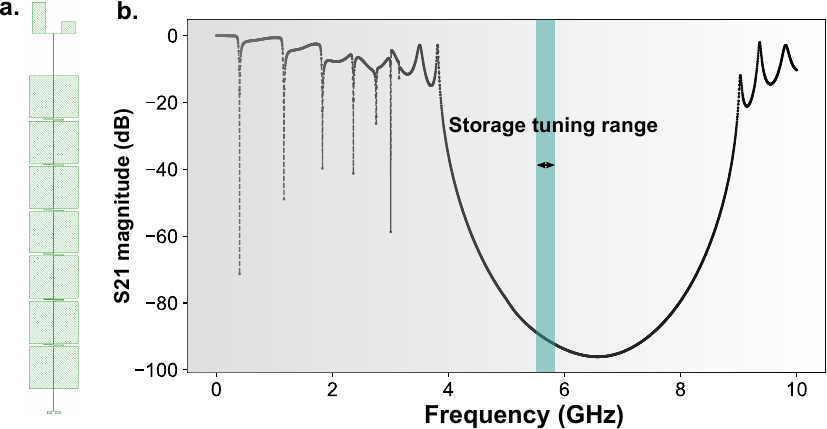}
    \caption{(a) SQUID coupler and on-chip flux line design. (b) Ansys HFSS simulated filter $S_{21}$ response. The tuning range of the storage mode is highlighted in the same plot. }
    \centering
    \label{fig:coupler}
\end{figure}

The SQUID coupler design is illustrated in Fig.~\ref{fig:coupler}(a). This design features two large metal pads on the chip's top side, which provide capacitive coupling to the storage mode. A coupling strength of approximately $g \sim 170$ MHz is extracted from Black-box quantization simulations. The flux loop of the coupler is integrated with a single on-chip low-pass filter. The flux line is configured as a Coplanar Strip Line (CPS) structure. The low-pass filter is comprised of seven sections, alternating between capacitor segments with large pads (offering low impedance $Z_{\text{lo}}$) and inductor segments with meander-shaped lines (providing high impedance $Z_{\text{hi}}$). This configuration establishes an Edge-coupled microstrip CPS filter. 

Fig.~\ref{fig:coupler}(b) presents the ANSYS HFSS simulation results of our CPS filter. The primary characteristics of our filter design are the stopband center, bandwidth, and depth. These parameters are crucial for protecting the coupler and storage mode coherence. The length of each filter section primarily determines the stopband center frequency. The lengths of the low and high-impedance sections are equal for optimal performance. The stopband bandwidth can be effectively widened by increasing the impedance $Z_{\text{hi}}/Z_{\text{lo}}$, which is around $22$ for our experiment. However, in our experimental setup, this ratio is constrained by the physical dimensions of the high-impedance sections: the metal width (2 $\mu$m), the gap between the capacitor pads (2 $\mu$m), and the overall size of the capacitor parts, which is bound by the chip width. Adding more filter sections increases the stopband depth, although the coupler chip length limits this.

\section{Calibration }

\begin{table}[t]

		\begin{tabular}{c c c}
		    \hline
		    \hline
		     Device Parameter & Symbol & Value \\     
		    \hline
      Transmon Frequency & $\omega_q$ & $2\pi\times 4.811$ GHz  \\ 
       Transmon anharmonicity & $\alpha_q$ & $-2\pi\times 176.7$ MHz \\ 
      Transmon decay time & $T_1^q$ & $34\pm 0.5\,\mu$s  \\
      Transmon Ramsey time & $T_2^q$ & $31\pm 0.6\,\mu$s  \\ 
      Transmon background photon & $\Bar{n}_q$ & $1.4\pm 0.2 \times 10^{-2}$  \\
            \hline
    Storage frequency & $\omega_s$ & $2\pi\times 5.694$ GHz \\
    Storage decay time & $T_1^s$ & $65 \pm 10\,\mu$s \\
    Storage Ramsey time & $T_2^s$ & $40\pm 10\,\mu$s \\
    Storage-Transmon Stark shift & $2\chi$ & $-2\pi\times 3.31$ MHz \\
    Storage background photon & $\Bar{n}_s$ & $8.6\pm 0.5\times 10^{-3}$ \\ \hline
    Readout frequency & $\omega_r$ & $2\pi\times 6.733$ GHz  \\ 
    Readout time & $t_r$ & $ 2.3\,\mu$s  \\
    Readout releasing time & $t_l$ & $ 5\,\mu$s \\  
    Reset + Integration time & $t_c$ & $ 500\,\mu$s \\ 
    Readout-Transmon Stark shift & $2\chi_{qr}$ & $-2\pi\times 0.28$ MHz \\
    Readout fidelity ($\ket{g}$) & $F_{gg}$ & $97.8\pm 1\%$  \\ 
    Readout fidelity ($\ket{e}$) & $F_{ee}$ & $93.8\pm 1\%$  \\ \hline \hline
		\end{tabular}
		\caption{Device parameters measured at $\Phi_{\text{ext}}=0$}
		\label{table:device}
\end{table}

\begin{figure*}[t]
    \centering
    \includegraphics[width=1.5\columnwidth]
    {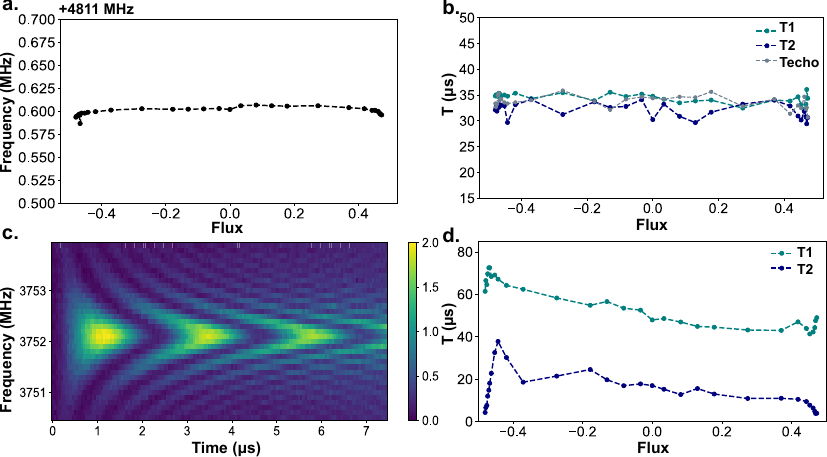}
    \caption{Transmon and storage mode calibration. Measured transmon frequency $\omega_q$ (a) and coherence (b) versus DC flux bias shows that our transmon can be approximated as a flux-independent element in our system. (c) shows the chevron plot of the sideband $\ket{f0} \leftrightarrow \ket{g1}$ between the transmon and the storage. The color bar indicates the photon population in the transmon after different frequencies and duration of the sideband drive. The storage mode coherences at different flux points are shown in (d). Marker sizes are larger than the error bars.}
    \centering
    \label{fig:calibration}
\end{figure*}

\begin{figure}[t]
    \centering
    \includegraphics[scale=2.1]{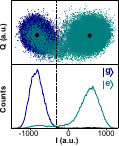}
    \caption{Single shot readout of transmon states. Transmon is prepared in one of its possible states $\{\ket{g}, \ket{e}\}$ in blue and cyan. $30000$ single shot data for each state are measured. The black dash line indicates the demarcation location, where the readout is assigned to $\ket{g}$ when the I channel signal is smaller than the demarcation line and assigned to $\ket{e}$ for the other case. Signals are rotated on the I-Q plane to maximize the contrast in the I channel.}
    \centering
    \label{fig:single_shot}
\end{figure}

To conduct the dark matter detection experiment, the following parameters are calibrated: transmon frequency $\omega_q$, single-shot readout fidelity, transmon-storage coupling $\chi(\Phi_{\text{ext}})$, storage frequency $\omega_s(\Phi_{\text{ext}})$, storage displacement drive, storage coherence $T^s_1$, and transmon and storage background photon populations $\{\Bar{n}_q, \Bar{n}_s\}$. For the flux-tunable parameters, such as $\omega_s(\Phi_{\text{ext}})$, we calibrate the values at $34$ separate flux points $\Phi_{\text{ext}}$ varying between $-0.4807$ and $0$. We perform a linear extrapolation fit for all other values to extract the numbers.

The transmon frequency $\omega_q$ is calibrated through the Rabi-Ramsey-Rabi sequence: Using the rough transmon frequency $\omega_q$ measured through two-tone spectroscopy, the first transmon Rabi drive signal calibrates the $\pi$ and $\pi/2$ pulse length. After that, the Ramsey experiments calibrate the transmon frequency, followed by the second transmon Rabi drive signal to fine-adjust the $\pi$ and $\pi/2$ pulse length. The measured transmon frequency, $T_1$, and $T_2$ at different flux points are shown in Fig.~\ref{fig:calibration} (a) and (b). The measured transmon single shot fidelity (flux-independent) is shown in Fig.~\ref{fig:single_shot}, where the transmon decay event during readout limits the fidelity.

The transmon-storage coupling strength $\chi$ is measured using Ramsey experiments. Initially, we prepare the transmon in the second excited state, $\ket{f}$, by applying two sequential single-transmon rotations. Following this, we drive the transmon port to activate the $\ket{f0}\leftrightarrow\ket{g1}$ sideband transition between the transmon and storage mode (Shown in Fig.~\ref{fig:calibration} (c)). This sideband prepares the state $\ket{1}$ in the storage mode. Subsequently, a Ramsey experiment is conducted on the transmon to determine its frequency. Compared to the storage mode in the $\ket{0}$ state, the observed frequency shift is $2\chi$. And the parity waiting time $t_p=\pi/|2\chi|$ is simultaneously calibrated. 

The storage frequency $\omega_s$ is calibrated using Ramsey experiments. For each DC flux point targeted for calibration, the transmon is initially prepared in the state $\frac{1}{\sqrt{2}}(\ket{g} + \ket{f})$. This is achieved through a $\pi/2$ rotation within the ${\ket{g}, \ket{e}}$ subspace followed by a $\pi$ rotation within the ${\ket{e}, \ket{f}}$ subspace. We then activate the $\ket{f0} \leftrightarrow \ket{g1}$ sideband transition to prepare the state $\frac{1}{\sqrt{2}}(\ket{0} + \ket{1})$ in the storage mode. By varying the waiting time before reapplying the $\ket{f0} \leftrightarrow \ket{g1}$ transition and decoding the storage state into the transmon, we fit the Ramsey fringe data to calibrate $\omega_s$ and the storage $T_2$.

After calibrating $\omega_s$, the storage coherence $T_1^s$ is measured. This involves preparing the storage mode in state $\ket{1}$ using the $\ket{f0} \leftrightarrow \ket{g1}$ sideband transition, waiting for a variable delay, then reapplying the $\ket{f0} \leftrightarrow \ket{g1}$ transition and measure the transmon population. The storage decay and Ramsey times at different flux points are documented in Fig.~\ref{fig:calibration} (d).

\begin{figure}[t]
    \centering
    \includegraphics[scale=1]{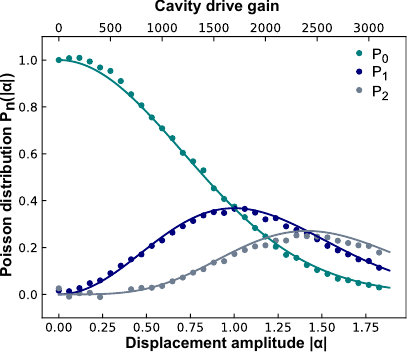}
    \caption{Mapping drive gain $\epsilon$ to storage displacement $\alpha=\epsilon/c$. Storage photon populations in $\{\ket{0}, \ket{1}, \ket{2}\}$ are measured after a constant $\SI{0.5}{\micro\second}$ storage displacement with a varied drive gain. The data are fitted to extract $c$.  Negative data points are neglected to fit the results better. Marker sizes are larger than the error bars.}
    \centering
    \label{fig:displacement}
\end{figure}

To calibrate the correspondence between
the drive gain and alpha, we displace the storage cavity using a $\SI{0.5}{\micro\second}$ square pulse, generating a coherent state $\ket{\alpha}$. The magnitude of $\alpha$ depends on the DAC channel's drive gain $\epsilon$. We establish the $\epsilon$-$\alpha$ mapping by varying $\epsilon$ and measuring the resultant displacement $\ket{\alpha}$, as illustrated in Fig.~\ref{fig:displacement}. The measurement of $\alpha$ involves applying a photon-number-resolved $\pi$ pulse to the transmon at frequencies $\omega_q-n\chi$ for various integers $n$. To improve accuracy, we swept the transmon frequency around $\omega_q-n\chi$ and fitted the signal peak amplitude to extract the cavity Fock state population at each $\alpha$. For small $\alpha$, higher cavity Fock states such as $\ket{2}$ have small populations, making the fitted signal peak amplitude unstable and occasionally negative. These points were therefore excluded to achieve a more reliable calibration of the $\epsilon$-$\alpha$ mapping. This resolved $\pi$ pulse maps the population of the Fock state $\ket{n}$ in the coherent state $\ket{\alpha}$ to the excited state $\ket{e}$ of the transmon. The relationship between the population $P_n$ of the Fock state $\ket{n}$ and $\alpha$ is given by: 

\begin{align}
P_n(\alpha=c\epsilon)&=\frac{\alpha^{2n}e^{-\alpha^2}}{n!}.
\end{align}
Here, $c$ represents the calibration constant that maps the drive gain to the amplitude of the coherent state. By measuring the populations of the Fock states ${\ket{0}, \ket{1}, \ket{2}}$ for various values of $\epsilon$ and fitting these populations, we extract the value of $c$ (See Fig.~\ref{fig:displacement}). 

\begin{figure}[t]
    \centering
    \includegraphics[scale=1]{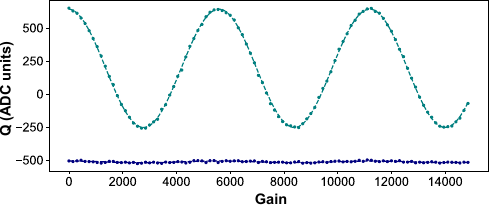}
    \caption{Transmon temperature measurement. Transmon $\ket{e}-\ket{f}$ amplitude Rabi signal with (green) and without (purple) initial $\ket{g}-\ket{e}$ rotations are plotted. Dash lines show the fitting results. }
    \centering
    \label{fig:qubit temperature}
\end{figure}

To characterize the transmon background population $\Bar{n}_q$, we measure the $\ket{e}-\ket{f}$ Rabi oscillation amplitude ratio with and without the initial transmon $\ket{g}-\ket{e}$ rotation (See Fig.~\ref{fig:qubit temperature}). This extracts the transmon population at the ground state $P(g)$ and at the excited state $P(e)$ in its steady state. Ignoring the transmon population above the first excited state, we attain $P(g)=0.985$ and $P(e)=0.015$, corresponding to a qubit temperature of $53$ mK.

Following the photon counting scheme in Fig. 2(a), two experimental data are shown in Fig.9 by gathering the single-shot data following N parity measurements (N = 25).

\begin{figure}[t]
    \centering
    \includegraphics[scale=1]{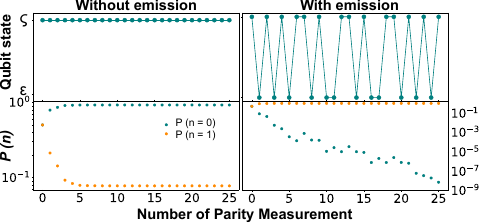}
    \caption{ Photon counting experiments. Two examples show the measured qubit readout sequences for a cavity initialized in $\ket{N} = 0$ Fock state followed by a small displacement drive. we obtain a series of transmon states for each experiment, which remains constant (left) when the storage mode is in the ground state and displays an oscillating pattern when the storage mode is excited
(right).}
    \centering
    \label{fig:qubit temperature}
\end{figure}

\section{Hidden Markov model}

\label{sec:hidden}

We model our storage-transmon-readout system for each storage tuning point as a hidden Markov process. The storage and transmon states are treated as 
hidden variables emitted as the readout signal~\cite{PhysRevLett.126.141302}. The transition matrix $T$ characterizes how the storage-transmon state evolves, and the confusion matrix $E$ determines the readout signal $\{g, e\}$ given any storage-transmon state:

\begin{align}
T&=\
\begin{blockarray}{ccccc}
\ket{0g} & \ket{0e} & \ket{1g} & \ket{1e} & \\
\begin{block}{(cccc)c}
P_{00}P_{gg}  & P_{00}P_{ge}    & P_{01}P_{ge}  & P_{01}P_{gg}  &\ \ket{0g}\\
P_{00}P_{eg}  & P_{00}P_{ee}   & P_{01}P_{ee}  & P_{0q}P_{eg}  &\ \ket{0e} \\
P_{10}P_{gg}  & P_{10}P_{ge}  & P_{11}P_{ge}  & P_{11}P_{gg}  &\ \ket{1g} \\
P_{10}P_{eg}  & P_{10}P_{ee} & P_{11}P_{ee} & P_{11}P_{eg} &\  \ket{1e}      \\
\end{block}
\end{blockarray}\\
E&= \
\begin{blockarray}{ccc}
g & e & \\
\begin{block}{(cc)c}
F_{gg}  & 1-F_{gg}     &\ \ket{0g}\\
1-F_{ee}  & F_{ee}     &\ \ket{0e} \\
F_{gg}  & 1-F_{gg}    &\ \ket{1g} \\
1-F_{ee}  & F_{ee}  &\  \ket{1e}    \\
\end{block}
\end{blockarray}
\end{align}

The transition matrix $T$ elements are calculated based on the coherence and background photon population in the transmon and storage~\cite{PhysRevLett.126.141302}:

\begin{align}
\left\{ \begin{array}{l}
    P_{gg} = 1 - P_{ge},\ P_{ee} = 1 - P_{eg}\\
    P_{ge} = \Bar{n}_q[1-e^{-t_m/{T_1^q}}]+1- e^{-t_p/T_2^q} \\
    P_{eg} = 1-e^{-t_m/{T_1^q}}+1- e^{-t_p/T_2^q} \\
    P_{00} = 1 - P_{01},\ P_{11} = 1 - P_{10}\\
    P_{01} = \Bar{n}_s[1-e^{-t_m/{T_1^s}}] \\
    P_{10} = 1-e^{-t_m/{T_1^s}}    
\end{array} \right.
\end{align}

Here $P_{gg}$, $P_{ee}$, $P_{00}$, and $P_{11}$ correspond to event where no error occurs. Heating and decay in transmon and storage during readout period $t_m$ and transmon dephasing during parity waiting time $t_p$ contribute to the state changes. Here $t_m=t_r+t_l=\SI{7.3}{\micro\second}$ is the full readout time. 

The confusion matrix $E$ is directly measured through the single shot experiments (See Fig.~\ref{fig:single_shot}). Here $F_{gg}$ and $F_{ee}$ correspond to the case when the measurement result matches the transmon state.

Following Reference~\cite{PhysRevLett.126.141302}, the initial storage state probabilities $P(i=0)$ and $P(i=1)$ are reconstructed through the backward algorithm after $N$ readout signals $(R_1,...,R_N)$: 

\begin{align}
P(i)&=E_{\ket{ig}}^{R_1}T_{\ket{ig}}^{a_1}E_{a_1}^{R_2}T_{a_2}^{a_3}...T_{a_{N-1}}^{a_N}E_{a_N}^{R_N} \nonumber \\
&+E_{\ket{ie}}^{R_1}T_{\ket{ie}}^{b_1}E_{b_1}^{R_2}T_{b_2}^{b_3}...T_{b_{N-1}}^{b_N}E_{b_N}^{R_N}
\end{align}

Here, $\{a_j, b_j\}$ are the indices that are summed over. The larger the ratio $\lambda=P(i=0)/P(i=1)$ is, the less likely the storage mode is excited. The threshold value $\lambda_{\text{thresh}}$ is chosen to determine if the storage is at $\ket{1}$. To characterize the detector's $\delta$ and $\eta$ at different flux tuning points, we inject a list of known weak displacements $\alpha$ into the storage mode and perform the parity measurement with $N=25$ and $\lambda_{\text{thresh}}=125$ in the Hidden Markov model analysis. We fit the injected photon population ($\Bar{n}_{inj}=|\alpha|^2$) and measured photon population ($\Bar{n}_{meas}$) to the equation:

\begin{align}
\Bar{n}_{inj} = \eta \times \Bar{n}_{meas} + \delta
\end{align}

\begin{figure}[t]
    \centering
    \includegraphics[width=\columnwidth]{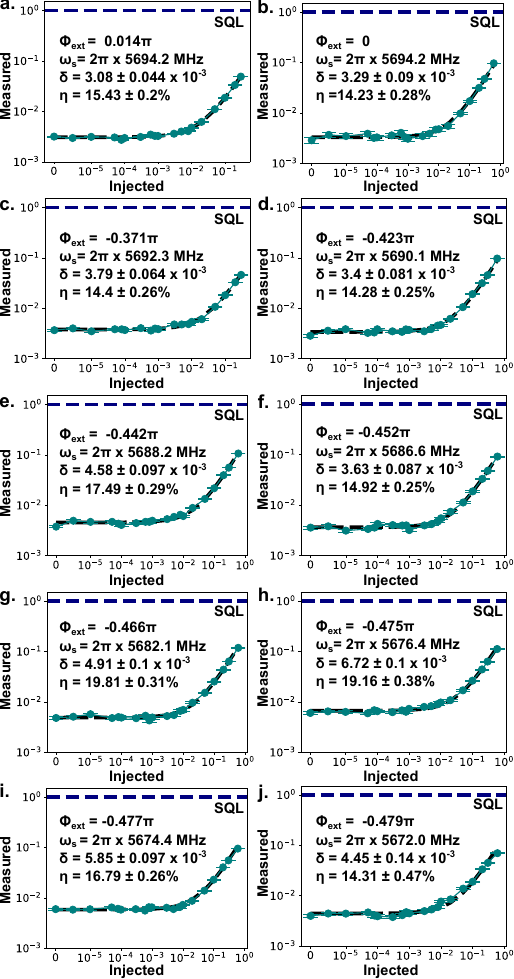}
    \caption{Detector efficiency characterization at different DC flux biasing points.}
    \centering
    \label{fig:character_flux}
\end{figure}

\begin{figure}[t]
    \centering
    \includegraphics[scale=1.1]{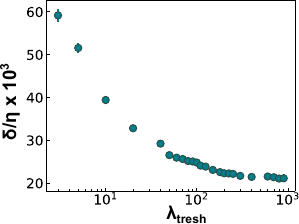}
    \caption{Efficiency corrected false-positive probability ($\Phi_{\text{ext}}=0$).}
    \centering
    \label{fig:threshold}
\end{figure}

\begin{figure}[t]
    \centering
    \includegraphics[scale=1.0]{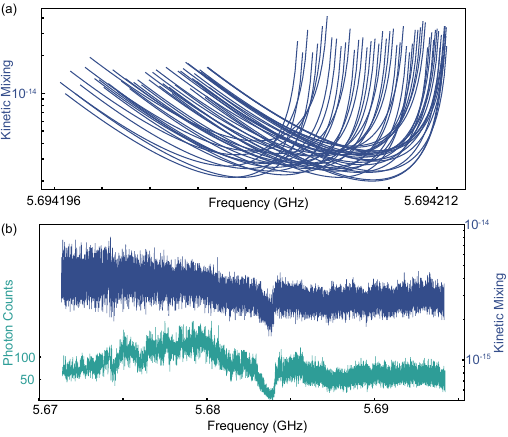}
    \caption{(a) A zoom in the our tuning range shows a family of curves describing the calculated exclusion for each cavity tuning. The minimum envelope of these curves describes the dark matter exclusion across our tuning range. (b) Positive photon counts and mixing angle exclusion data. The mixing angle exclusion data has applied the minimum envelop.}
    \centering
    \label{fig:counts_mixing}
\end{figure}

The results at different $\Phi_{\text{ext}}$ (in addition to the data presented in the main text) are shown in Fig.~\ref{fig:character_flux}. Here $\lambda_{\text{thresh}}=125$ are chosen based on the efficiency corrected false
positive probability $\delta/\eta$ (See Fig.~\ref{fig:threshold}): Increasing $\lambda_{\text{thresh}}$ will exponentially suppress the detector's false positive probability $\delta$ while reducing the detector efficiency $\eta$. The ratio $\delta/\eta$ continues to decrease and become almost insensitive to $\lambda_{\text{thresh}}$ when $\lambda_{\text{thresh}}>100$. For a large enough threshold $1/(\lambda_{\text{thresh}}+1)<\delta$, the qubit readout error is no longer the dominant error source and smaller $\lambda_{\text{thresh}}$ helps speed up the detection. In our case, $1/(\lambda_{\text{thresh}}+1)<\delta$ is satisfied when $\lambda_{\text{thresh}}=125$. Fig.~\ref{fig:counts_mixing} shows the positive photon counts and mixing angle exclusion data.

The hidden Markov model counts on independent measurements of the probabilities involved in the transition and emission matrices. The elements of these matrices depend on the parameters of the experiment and the device, including the lifetime and dephasing times of the qubit and cavity, qubit population, and readout fidelities. In our dark-photon detection experiments, the detector efficiencies are linearly extrapolated at other $\Phi_{\text{ext}}$ points.

\section{Photon detector efficiency simulation}
\label{app:sim}

\begin{figure}[t]
    \centering
    \includegraphics[scale=1]{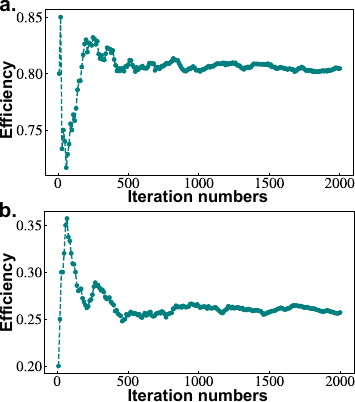}
    \caption{Simulated detector efficiency without (a) and with (b) decoherence. We use the device parameter experimentally measured at DC flux=0 (Table.~\ref{table:device}) in the master equation simulation. The fast transmon $\pi/2$ rotation pulse used in both simulation and experiment is a Gaussian pulse with a $2\sigma$ cutoff ($\sigma=10$ ns).}
    \centering
    \label{fig:eff_simulation}
\end{figure}

To understand the limiting factor of the tunable photon detector efficiency, we perform the master equation simulation using QuTip. We focus on the storage-transmon subsystem that implements the QND measurement of storage excitation and ignore the other components for simplicity. All simulations are carried out in a Hamiltonian of dimension $3\times4$ (3 levels for the transmon and 4 levels for the storage mode). The Hamiltonian used in the simulation is:
\begin{align}
H_{qs} & = \omega_{q}a_{q}^{\dagger}a_{q}+\alpha a_{q}^{\dagger}a_{q}^{\dagger}a_{q}a_{q}/2 \nonumber \\
& + \omega_{r}a_{r}^{\dagger}a_{r}+2\chi a_{q}^{\dagger}a_{q}a_{r}^{\dagger}a_{r}
\end{align}

Since both transmon and storage modes are only actively used up to the first excited state in the experiment, we approximate the transmon as a Kerr-oscillator. The simulation starts with the first excited state $\ket{1}$ in the storage mode and the ground state $\ket{g}$ for the transmon, followed by a sequence of 25 parity check measurements. Each parity check contains two $\pi/2$ transmon rotations ($2\sigma$ cutoff Gaussian pulse, $\sigma=$10 ns) gapped by a waiting time $\frac{\pi}{2\chi}$, and a following waiting time of $7.3\,\mu$s that represents the total readout time (including readout photon releasing time). To simulate the transmon readout error, we first calculate the ground state probability of the transmon $P_{0}$ and assign all the other transmon states' probability to the first excited state $P_{1}=1-P_{0}$. The transmon state probability vector $\vec{P}=\left(P_{0},P_{1}\right)^{T}$ is further multiplied by the readout confusion matrix $\bf F$ measured in the experiment:$\vec{\tilde{P}}={\bf F} \vec{P}$. After each parity measurement, the transmon state is assigned based on the $\vec{\tilde{P}}$ probability. A collection of 25 measured transmon states after the parity check sequence is further analyzed using the Markov model analysis discussed before. We choose the same threshold $\lambda_{thresh}=125$ as used in the experiment to determine the state of the cavity. The whole simulations are repeated 2000 times, and the average time measuring $\ket{1}$ for the cavity state is the detector efficiency.    

We compare the simulated detector efficiency without and with decoherence and plot the results in Fig.~\ref{fig:eff_simulation}. The relatively slow transmon rabi pulse makes the transmon rotation dependent on the cavity state, which results in an efficiency of $80\%$ even without decoherence. The room temperature control hardware power limits this and can be easily fixed by adding extra amplifiers. The decoherence in the system, especially the $T_1$ of the cavity, limits the detector efficiency to $25\%$. Better cavity etching, sealing, and a modified on-chip filter design will help push the efficiency close to the decoherence-free limit.

\section{Converting cavity background counts to hidden-photon exclusion}

\begin{table}[t]

		\begin{tabular}{c c c}
		    \hline
		    \hline
		     Parameter & $\Theta$ & $\sigma_{\Theta}$ \\     
		    \hline
      Detector efficiency $\eta$ ($\%$) & $\left[6.81, 19.81\right]$ & $2.1$ \\
      Storage quality factor $Q_s$ ($\times 10^6$) & $\left[2.15, 2.98\right]$ & $0.009$  \\ 
      Storage frequency $\omega_s$ (MHz) & $\left[5671.4, 5694.2\right]$ & $0.1$  \\ 
      Storage volume $V$ (cm$^3$) & $6.452$ & $0.175$  \\ 
      Storage form factor $G$ & $0.219$ & $0.003$  \\ \hline \hline
		\end{tabular}
		\caption{Mean $\Theta$ and standard deviation $\sigma_{\Theta}$ for experimental parameters. The standard deviation for $\eta$ is extracted from the injected and measured photon fitting with $\lambda_{\text{thresh}}=125$. The standard deviation in $\omega_s$ is calculated based on the storage $T_2$. The storage volume standard deviation is calculated assuming each dimension's machining tolerance of $0.005$ inches. The tunable parameters' ranges are listed in the table.  We use the largest value for all tunable parameters' standard deviation in the analysis as the lower bound performance of our detector. }
		\label{table:cavity parameter}
\end{table}

For a dark matter candidate on resonance with the storage frequency ($m_{DM}c^2=\hslash\omega_c$), the rate of photons deposited in the storage cavity by the coherent build-up of the electric field in one storage cavity coherence time is given by~\cite{PhysRevD.92.075012, PhysRevLett.126.141302}:

\begin{align}
\frac{{dN_{HP}}}{dt}=\frac{U/\omega_s}{T_1^s}=\frac{2E^2V}{\omega_s}\frac{\omega_s}{Q_s}=\frac{{J_{DM}^2}{Q_{DM}}GV}{2m^2}
\end{align}

The storage volume $V$ is $3.556\times3.81\times0.476\,
 \text{cm}^3=6.452\,\text{cm}^3$. Here, we assume the storage mode quality factor $Q_s$ is larger than the dark matter quality factor $Q_{DM}$, such that the storage field displacement from dark photons can be approximated as a random walk. The geometric form factor $G$ for our rectangular cavity used in the experiment is~\cite{PhysRevLett.126.141302}:
\begin{align}
G=\frac{1}{3}\frac{\mid{\int}dV{E_z}\mid^2}{V{\int}dV\mid{E_z}\mid^2}=\frac{1}{3}\frac{2^6}{\pi^4}
\end{align}

The hidden-photon generated current is set by the density of dark matter in the galaxy $\rho_{DM}=0.45\,GeV/cm^3=2\pi\times9.67\times10^{19}\,GHz/cm^3$~\cite{PhysRevD.104.095029}:

\begin{align}
J_{DM}^2=2\varepsilon^2{m^4}{A^{'2}}=2\varepsilon^2{m^2}\rho_{DM}
\end{align}

The photon deposition rate now becomes:
\begin{align}
\frac{dN_{HP}}{dt}=\varepsilon^2\rho_{DM}Q_{DM}GV
\end{align}

The photon rate and the integration time ($T_I(\Phi_{\text{ext}})=T_1^sN_{meas}\in\left[1.29s, 1.38s\right]$ ) determines the total number of photons expected to be deposited in the storage cavity before each measurement: 

\begin{align}
N_{HP}=\frac{dN_{HP}}{dt}\times T_I(\Phi_{\text{ext}})=\varepsilon^2\rho_{DM}Q_{DM}GVT_1^sN_{meas}
\end{align}

By performing single-photon detection at each DC flux point, we perform a hidden-photon search across the storage tuning band. In our experiments, we perform $N_{\text{meas}}=20,000$ measurements at each tuning point, and the observed counts $N_{\text{obs}}$ are shown in supplementary text Figure  ~\ref{fig:smooth_background}. To convert the counted photons into an exclusion on the hidden-photon mixing angle, we first smooth the measured data to obtain a background of photon counts as a function of the cavity frequency. The smoothing is done by applying a Savitzky–Golay filter with polynomial order 4 and window length of 112, approximately the square root of the number of frequency bins 12695 (See Figure~\ref{fig:smooth_background}). This yields a photon background $N_{back}$

\begin{figure}[t]
    \centering
    \includegraphics[scale=0.7]{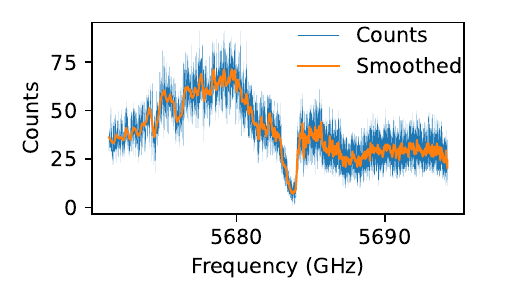}
    \caption{Background photon counts as a function of cavity frequency and the smoothed background for the purposes of background subtraction.}
    \centering
    \label{fig:smooth_background}
\end{figure}

At each frequency, we calculate the 95\% exclusion for the hidden-photon coupling parameter $\varepsilon$, which includes marginalizing over experimental parameter variation (See Table~\ref{table:cavity parameter}). Additionally at some frequency points we encounter the situation where the background estimated counts are much higher than the observed counts. This leads the unphysical conclusion that there is no hidden-photon candidate that could produce the observed signal. We follow the procedure outlined in \cite{Read2000} and normalize the background + signal hypothesis cumulative distribution by the background only hypothesis distribution.
At each frequency, we calculate the confidence level for a given hidden-photon coupling parameter $\varepsilon$, which includes marginalizing over experimental parameter variation as shown in Equation~\ref{eqn:prior_update}. 

\begin{align}
\label{eqn:prior_update}
U(\omega, \varepsilon)  &= \int \prod_i d\Theta_i' F(\Theta_i - \Theta_i', \sigma_{\Theta_i}) \nonumber\\
&\left[ 1-\frac{\sum_0^{N_{obs}} \frac{(N_{back} + N_{test})^{N} e^{-(N_{back} + N_{test})}}{N!}}{\sum_0^{N_{obs}} \frac{(N_{back})^{N} e^{-(N_{back})}}{N!}} \right]\\
% e^{-\sigma_{test}^2/2 + \sigma_{test} \sigma_{obs}} \\
% \sigma_{obs} &= \frac{N_{obs} - N_{back}}{\sqrt{N_{back}}} \\
N_{test} &= \frac{\eta \varepsilon^2 \rho_{DM} Q_{DM} Q_c G V_c N_{meas}}{ \omega} \nonumber\\
F(\mu, \sigma) &= \frac{e^{-\mu^2/2\sigma^2}}{\sqrt{2\pi}\sigma} \nonumber
\end{align}

The parameters used to calculate the expected photon counts from a given dark matter candidate are measured and their uncertainties must be included in the analysis. Table~\ref{table:cavity parameter} reports the uncertainties of all nuisance parameters considered in the analysis. Uncertainties in the transmon frequency are not critical, as they are encompassed within the detector efficiency uncertainty. When the transmon frequency drifts, the corresponding rotation pulses ($\pi$ and $\pi/2$) will have slightly altered gate fidelities. This leads to minor variations in single-shot fidelity and thus in the detector efficiency. We weight these parameters each with a Gaussian distribution, $F(\mu, \sigma)$, and marginalize over them.

For a given frequency bin, we find the hidden-photon parameter $\varepsilon$ below which the observed photon count is produced with more that 5\% probability. Hidden-photon candidates which would produce the observed counts with less than 5\% probability are excluded at the 95\% confidence level. Additionally, we account for the lineshape of the dark matter \cite{PhysRevD.97.123006} by convolving the dark matter distribution with the exclusion from each cavity tuning. This yields a family of curves exhibiting the dark matter lineshape with maximum sensitivity when the dark matter mass is equal to the cavity frequency, see Figure ~\ref{fig:counts_mixing}. We now excluded dark matter candidates continuously across our tuning range as the minimum envelope of the family of curves coming from the above described procedure.

\section{Microwave measurement setup: cryogenic and room temperature wiring}

\begin{figure}[t]
    \centering
    \includegraphics[width=\columnwidth]{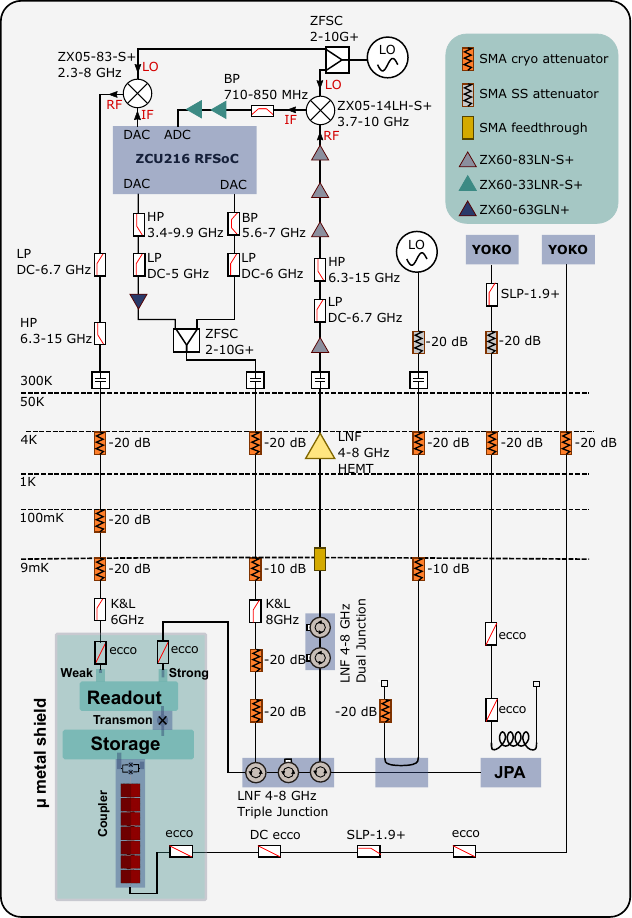}
    \caption{Wiring diagram inside the dilution refrigerator and the room temperature measurement setup.}
    \centering
    \label{fig:wire-diagram}
\end{figure}

Fig.~\ref{fig:wire-diagram} shows the room and cryogenic temperature measurement chain. The device is mounted on the mixing chamber plate of the dilution fridge with a 15 mK base temperature. Two layers of $\mu$ -metal are used to shield the external magnetic fields. All RF input signals are generated through the ZCU216 RFSoC board and attenuated and thermalized at each temperature stage of the dilution fridge. Two DC sources (Yokogawa GS200) are used to bias the DC flux of the coupler and the Josephson Parametric Amplifiers (JPA). One Signal Core SC5511A is used as the LO and split into two tones, one used for frequency up-conversion and sent to the device's weakly coupled port as the readout drive pulse; the other tone is used for frequency down-conversion of the readout signal and collected by the RFSoC ADC. The qubit and storage pulses are synthesized directly through different RFSoC DAC channels, combined at room temperature, and sent to the device's strongly coupled port. The readout signal goes through three circulators, then amplified by a JPA with 20dB gain, followed by two circulators and amplified with one LNF High-Electron-Mobility Transistor (HEMT) amplifier. One additional Signal Core SC5511A provides the JPA's 3-wave mixing modulation tones. DC flux tuning line is connected to the coupler chip through an SMA thread and wirebonds. All control lines contain SMA cryo attenuators at the base state, and quantum microwave eccosorbs in the $\mu$-metal can.

\clearpage

\bibliography{2Q}